\documentclass[12pt,aps,showpacs]{revtex4}%

\usepackage{amsmath}
\usepackage{amssymb}
\usepackage{dsfont} 

\usepackage{graphicx}
\usepackage{epsfig}
\usepackage{graphics}
\usepackage{epstopdf}

\usepackage{color}

\newcommand*{\cofrac}[2]{%
  {%
    \rlap{$\dfrac{1}{\phantom{#1}}$}%
    \genfrac{}{}{0pt}{0}{}{#1+#2}%
  }%
}

\newcommand{\be}{\begin{equation}}
\newcommand{\ee}{\end{equation}}
\newcommand{\bel}[1]{\begin{equation}\label{#1}}
\newcommand{\bea}{\begin{eqnarray}}
\newcommand{\eea}{\end{eqnarray}}
\newcommand{\balign}{\begin{align}}
\newcommand{\ealign}{\end{align}}
\newcommand{\ba}{\begin{array}}
\newcommand{\ea}{\end{array}}
\newcommand{\bfig}{\begin{figure}}
\newcommand{\efig}{\end{figure}}

\newcommand{\eref}[1]{(\ref{#1})}

\allowdisplaybreaks

\newcommand{\exval}[1]{\mbox{$\langle \, {#1}\, \rangle$}}

\newcommand{\matrixi}[1]{{\mathbf{#1}}}

\newcommand{\rmd}{\mathrm{d}}
\newcommand{\rme}{\mathrm{e}}

\newcommand{\sign}{\mathrm{sgn}}

\newcommand{\pdt}{\frac{\partial}{\partial t}}
\newcommand{\pdx}{\frac{\partial}{\partial x}}
\newcommand{\half}{\frac{1}{2}}

\newcommand{\I}{{\mathbb I}}

\newtheorem{theo}{Theorem}[section]

\newtheorem{rem}[theo]{Remark}
\def\qed{\hfill$\Box$\par\medskip\par\relax}

\begin{document}

\title{Exact scaling solution of the mode coupling equations
for non-linear fluctuating hydrodynamics in one dimension}
\author{V. Popkov$^{1}$, A. Schadschneider$^{2}$, J. Schmidt$^{2}$, 
G.M. Sch\"utz$^{3}$}
\affiliation{$^{1}$Helmholtz-Institut f\"ur Strahlen-und Kernphysik, 
Universit\"at Bonn, Nussallee 14-16, 53119 Bonn, Germany}
\affiliation{$^{2}$Institut f\"{u}r Theoretische Physik, Universit\"{a}t 
zu K\"{o}ln, Z\"ulpicher Str. 77, 50937 Cologne, Germany}
\affiliation{$^{3}$Institute of Complex Systems II, Theoretical Soft Matter 
and Biophysics, Forschungszentrum J\"ulich, 52425 J\"ulich, Germany}

\begin{abstract}
  We obtain the exact solution of the one-loop mode-coupling equations
  for the dynamical structure function in the framework of non-linear
  fluctuating hydrodynamics in one space dimension for the strictly
  hyperbolic case where all characteristic velocities are different.
  All solutions are characterized by dynamical exponents which are
  Kepler ratios of consecutive Fibonacci numbers, which includes
  the golden mean as a limiting case. The scaling form of all
  higher Fibonacci modes are asymmetric L\'evy-distributions.
  Thus a hierarchy of new dynamical universality classes is
  established. We also compute the precise numerical value
  of the Pr\"ahofer-Spohn scaling constant to which  
  scaling functions obtained from mode coupling theory are sensitive.
\end{abstract}

\maketitle

\newpage 

\section{Introduction}

Recently new insights into the dynamical universality classes of
nonequilibrium systems have been gained.
In the presence of slow modes due to locally conserved currents
(such as energy, momentum, density etc.) not only the well-established 
diffusive and the Kardar-Parisi-Zhang (KPZ) universality classes arise in one space 
dimension, but also a heat mode and other long-lived modes with unexpected
scaling properties were discovered \cite{vanB12,Popk15a,Spoh15}. 
Going further we have demonstrated in \cite{Popk15b} that  in the presence of
several conserved quantities there is an infinite
family of dynamical universality classes that is
characterized by dynamical exponents which take the form of Kepler
ratios $z_i=F_{i+2}/F_{i+1}$ where $F_i = 1,1,2,3,5,\dots$ are the
Fibonacci numbers defined recursively by $F_{i} = F_{i-1} + F_{i-2}$
with starting values $F_1=F_2=1$. This conclusion was based on a
scaling analysis of the mode coupling equations for non-linear
fluctuating hydrodynamics (NLFH) \cite{Spoh14} and supported by
extensive Monte-Carlo simulations of multi-lane asymmetric exclusion
processes.  The first level of the hierarchy (apart from the usual
diffusion with $z=2=z_1$) includes the Kardar-Parisi-Zhang
(KPZ) universality class with $z=3/2=z_2$ which continues to
inspire both due to its links to intriguing mathematical problems and
beautiful experimental results, see e.g. the special issue J. Stat.
Phys. 160, (2015) dedicated to it, and in particular the review by
Halpin-Healey and Takeuchi \cite{Halp15}.  Also the golden mean, which
is the limiting Kepler ratio $z_\infty=\varphi\approx 1.618...$,
occurs in systems with at least two conservation laws
\cite{Popk15a,Spoh15}.

NLFH has emerged as a universal tool to analyze general
one-dimensional systems such Hamiltonian dynamics \cite{vanB12,Roy15},
anharmonic chains \cite{Spoh14,Spoh15,Delf07,Poli11,Mend13,Das14} or
driven diffusive systems
\cite{Popk15b,Popk15a,Das01,Naga05,Ferr13,Popk14,Chak16}. The theory
is robust. The essential ingredients appear to be only the 
above-mentioned locally conserved currents and
long-time dynamics dominated by the long wave length modes of the
associated conserved quantities. 
Mathematically rigorous results for some specific models
support the validity of the theory \cite{Bern14,Komo16}.
It is the purpose of this work to provide a detailed analysis of the
one-loop mode-coupling equations for the dynamical structure function
for an arbitrary number of conservation laws in the strictly hyperbolic 
setting where all characteristic velocities are different.

\section{Computation of the dynamical structure function}

\subsection{Basis of nonlinear fluctuating hydrodynamics}

Consider an interacting system with $n$ locally conserved currents
$j_\lambda$ associated to physical quantities such as energy, momentum,
particle numbers etc. that are conserved under the microscopic dynamics
of the system. The starting point for investigating the
large-scale dynamics is the system of conservation laws
\bel{hydro}
\pdt \vec{\rho}(x,t) + \pdx \vec{j}(x,t) = 0
\ee
where component $\rho_\lambda(x,t)$ of the vector $\vec{\rho}(x,t)$ is
a coarse-grained conserved quantity and the component $j_\lambda(x,t)$
of the current vector $\vec{j}(x,t)$ is the associated locally
conserved current.  We shall refer to the $\rho_\lambda(x,t)$ as
densities.  Notice that in our convention $\vec{\rho}$ and $\vec{j}$
are regarded as column vectors. Transposition is denoted by a
superscript $T$.

This system of conservation laws can be obtained from the law of a
large numbers and the postulate of local equilibrium
\cite{Spoh91,Kipn99}. Thus the current is a function of $x$ and $t$
only through its dependence on the local conserved densities.  Hence
these equations can be rewritten as
\begin{equation}
\label{hyper}
\frac{\partial}{\partial t} \vec{\rho}(x,t) +
\bar{\matrixi{J}} \frac{\partial}{\partial x} \vec{\rho}(x,t) = 0
\end{equation}
where $\bar{\matrixi{J}} \equiv  \bar{\matrixi{J}}(\vec{\rho}(x,t))$ 
is the current Jacobian with matrix elements 
$\bar{J}_{\lambda\mu} = \partial j_\lambda / \partial \rho_\mu$.

To get some basic insight consider we first notice that constant
densities $\rho_\lambda$ are a (trivial) stationary solution of
\eref{hyper}.  Stationary fluctuations of the conserved quantities are
captured in the covariance matrix $\matrixi{K}$ of the conserved quantities
that we shall not describe explicitly. However, we have in mind the generic case 
where $\matrixi{K}$ is positive definite, i.e., we do not allow for vanishing
fluctuations of a locally conserved quantity that can occur in systems
with slowly decaying stationary correlations. We shall refer to
$\matrixi{K}$ as compressibility matrix.

Expanding the local densities $\rho_\lambda(x,t) = \rho_\lambda +
u_\lambda(x,t) $ around their long-time stationary values
$\rho_\lambda$ and taking a linear approximation (where $\bar{\matrixi{J}}$
is a constant matrix $\matrixi{J} \equiv \matrixi{J}(\vec{\rho})$ with matrix elements
determined by the stationary densities $\vec{\rho}$) leads to a system of coupled
linear PDE's which is solved by diagonalizing  $\matrixi{J}$. One
transforms to normal modes $\vec{\phi} =\matrixi{R} \vec{u}$ where
$\matrixi{R} \matrixi{J}  \matrixi{R}^{-1} = \mathrm{diag}(v_\alpha)$
and the transformation matrix $R$ is normalized such that $\matrixi{R}
\matrixi{K}\matrixi{R}^T =\mathds{1}$.  Thus one finds decoupled
equations $\partial_t \phi_\alpha = v_\alpha \partial_x \phi_\alpha$
whose solutions are travelling waves $\phi_\alpha(x,t) =
\phi^0_\alpha(x-v_\alpha t)$ with initial data $\phi_\alpha(x,0) =
\phi^0_\alpha(x)$. This shows that the eigenvalues $v_\alpha$ of  $\matrixi{J}$
play the role of characteristic speeds.

The product $\matrixi{J}\matrixi{K}$ of the Jacobian with the
compressibility matrix $\matrixi{K}$ is symmetric which can be proved
already on microscopic level \cite{Gris11} for sufficiently fast
decaying stationary correlations. This guarantees that on macroscopic
scale the full non-linear system (\ref{hyper}) is hyperbolic
\cite{Toth03}, i.e., all eigenvalues $v_\alpha$ of $\matrixi{J}$ are
guaranteed to be real.  If the eigenvalues $v_\alpha$ are
non-degenerate the system is called strictly hyperbolic. The
occurrence of complex eigenvalues signals macroscopic phase separation
\cite{Chak16}, consistent with the absence of fast decaying stationary
correlations on microscopic level, and coarsening dynamics.

Notice that \eref{hyper} is completely deterministic. In the NLFH
approach \cite{Spoh14} the effect of fluctuations is captured by adding a
phenomenological diffusion matrix $D$ and white noise terms $\xi_i$.
This turns \eref{hyper} into a non-linear stochastic PDE.
From renormalization group considerations it is known that polynomial
non-linearities of order higher than 4 are irrelevant for the large-scale
behaviour and order 3 leads at most to logarithmic corrections if the
generic quadratic non-linearity is absent \cite{Devi92}. Thus one expands
$\bar{\matrixi{J}}$ around the stationary densities $\vec{\rho}$ but keeps
only quadratic non-linearities  so that
the fluctuation fields $u_\lambda(x,t)$ satisfy the system of coupled
noisy Burgers equations
\begin{equation}
\label{coupledBurgers}
\partial_t \vec{u} =  - \partial_x \left( J_0 \vec{u} + \frac{1}{2} 
\vec{u}^T \vec{H} \vec{u} - D \partial_x  \vec{u}
+ B \vec{\xi} \right)
\end{equation}
where $\vec{H}$ is a column vector whose entries $(\vec{H})_\lambda
=\matrixi{H}^{\lambda}$ are the Hessians with matrix
elements $H^{\lambda}_{\mu\nu} = \partial^2 j_\lambda /(\partial \rho_\mu 
\partial \rho_\nu)$.
If the quadratic non-linearity is absent one has diffusive behaviour.

Using normal modes one thus arrives at
\begin{equation}
\label{normalmodes}
\partial_t \phi_\alpha = -  \partial_x \left( v_\alpha \phi_\alpha +
\vec{\phi}^T G^{\alpha} \vec{\phi} - \partial_x  (\tilde{D} \vec{\phi})_\alpha
+ (\tilde{B} \vec{\xi})_\alpha \right)
\end{equation}
with $\tilde{\matrixi{D}}=\matrixi{R}\matrixi{D}\matrixi{R}^{-1}$ and
$\tilde{\matrixi{B}}=\matrixi{R}\matrixi{B}$. The matrices
\begin{equation}
  \matrixi{G}^{\alpha} = \frac{1}{2} \sum_\lambda R_{\alpha\lambda}
  (\matrixi{R}^{-1})^T \matrixi{H}^{\lambda} \matrixi{R}^{-1}
\end{equation}
are the mode coupling matrices with the mode-coupling coefficients
$G^{\alpha}_{\beta\gamma}=G^{\alpha}_{\gamma\beta}$ which are,
by construction, symmetric. From the linear theory one concludes
that the fluctuation fields are peaked around $x_\alpha(t) =
x_\alpha(0) + v_\alpha t$. For short-range interactions fluctuations spread
generally sub-ballistically and therefore the width of the peak grows in
sublinearly time, as indeed will be seen explicitly below.

We stress that the macroscopic current-density relation given by the
components of the current vector $\vec{j}$ arises from the microscopic
model from the stationary current-density relation 
$\vec{j}(\vec{\rho})$. Similarly, the compressibility matrix
$\matrixi{K}$ is computed from the stationary distribution of the
microscopic model.  Hence the mode coupling matrices (and with them
the dynamical universality classes as shown below) are completely
determined by these two macroscopic stationary properties of the
system. However, the {\it exact} stationary current-density relations
and the {\it exact} stationary compressibilities are required.

The main quantity of interest are the dynamical structure functions
\bel{S-matrix}
S^{\alpha\beta}(x,t) = \exval{\phi^{\alpha}(x,t)\phi^{\beta}(0,0)}
\ee
(where $\exval{\ldots}$ denotes the stationary ensemble average)
which describe the stationary space-time fluctuations. 
Since we work
with normal modes we have the normalization
\bel{sfnorm}
\int_{-\infty}^\infty \rmd x \, S^{\alpha\beta}(x,t) = \delta_{\alpha,\beta}.
\ee
For strictly hyperbolic systems the characteristic velocities are all
different. As a result the off-diagonal elements of $S$ decay quickly
and for long times
and large distances one is left with the diagonal elements
which we denote by
\bel{S-diag}
S_\alpha(x,t):=S^{\alpha\alpha}(x,t).
\ee

The large scale behaviour of the diagonal elements is expected to 
have the scaling form
\bel{scalingform}
S_\alpha(x,t) \sim t^{-1/z_\alpha} f_{\alpha} (\xi_\alpha)
\ee
with the scaling variable
\bel{scvxi}
\xi_\alpha = (x-v_\alpha t) \, t^{-1/z_\alpha}
\ee
and dynamical exponent $z_\alpha$ which has to be determined and which
indicates the dynamical universality class of the mode $\alpha$. The exponent
in the power law prefactor follows from the conservation law.
In momentum space, with the Fourier transform convention
\bel{Def:FT}
\hat{S}_\alpha(k,t) := \frac{1}{\sqrt{2\pi}}\int_{-\infty}^\infty \rmd x\,
\rme^{- i k x} S_\alpha(x,t)
\ee
one has the scaling form
\bel{scalingformFT2}
\hat{S}_\alpha(k,t) \sim
\rme^{- iv_\alpha kt} \hat{f}_{\alpha} (k t^{1/z_{\alpha}})
\ee
where $\hat{f}_{\alpha}$ is the Fourier transform of the scaling function
\eref{scalingform}.


\subsection{Mode coupling equations}

The starting point for computing the diagonal elements
of the dynamical structure function are the mode coupling equations
\cite{Spoh14}
\be
\label{ModecouplingRS}
\partial_t S_{\alpha}(x,t) =  D_\alpha S_{\alpha}(x,t)
 +  \int_0^t \rmd s \int_{-\infty}^\infty \rmd y\,
 S_{\alpha}(x-y,t-s) M_{\alpha\alpha}(y,s)
\ee
with diffusion operator
\bel{DiffusionRS}
D_\alpha = -v_\alpha \partial_x + D_{\alpha} \partial_x^2
\ee
and memory term
\bel{MemoryRS}
M_{\alpha\alpha}(y,s) =
2 \sum_{\beta,\gamma} (G^{\alpha}_{\beta\gamma})^2
\partial_y^2 S_\beta(y,s)S_\gamma(y,s).
\ee
Only the diagonal elements $D_{\alpha}:= D_{\alpha\alpha}$ of the diffusion
matrix and of the memory kernel are kept here.

In momentum space this reads
\be
\label{ModecouplingFT}
\partial_t\hat{S}_\alpha(k,t) = - \widehat{D}_\alpha(k) 
\hat{S}_\alpha(k,t) -  \int_0^t \rmd s\, \hat{S}_\alpha(k,t-s)
\widehat{M}_{\alpha\alpha}(k,s)
\ee
with
\bel{DiffusionFT}
\widehat{D}_\alpha(k) = i v_\alpha k + D_{\alpha} k^2
\ee
and
\bel{MemoryFT}
\widehat{M}_{\alpha\alpha}(k,s) =
2 \sum_{\beta,\gamma} (G^{\alpha}_{\beta\gamma})^2
k^2 \int_{-\infty}^\infty \rmd q\, \hat{S}_\beta(q,s) \hat{S}_\gamma(k-q,s).
\ee

Finally we perform the Laplace transformation
\bel{Def:LT}
\tilde{S}_\alpha(k,\omega) := \int_0^\infty \rmd t\,
\rme^{- \omega t} \hat{S}_\alpha(k,t)
\ee
by multiplying \eref{ModecouplingFT}
on both sides by $\rme^{-\omega t}$ and integrating over $t$. This yields
\bel{ModecouplingFLT}
\tilde{S}_\alpha(k,\omega) = \frac{\hat{S}_\alpha(k,0)}{\omega + \widehat{D}_\alpha(k)
+  \tilde{C}_{\alpha\alpha}(k,\omega) }
\ee
with memory kernel
\bel{memoryFLT}
\tilde{C}_{\alpha\alpha}(k,\omega) = 2 \sum_{\beta,\gamma} 
(G^{\alpha}_{\beta\gamma})^2 k^2 \int_{0}^\infty \rmd s\, \rme^{-\omega s}
\int_{-\infty}^\infty \rmd q\,\hat{S}_\beta(q,s) \hat{S}_\gamma(k-q,s).
\ee
and $\hat{S}_\alpha(k,0)= 1/\sqrt{2\pi}$.

\begin{rem}
For $k=0$ the solution is trivial, with the exact result
$\hat{S}_\alpha(0,t) =1/\sqrt{2\pi}$ given by the Fourier
convention \eref{Def:FT} and the normalization \eref{sfnorm}.
\end{rem}

So far this is an exact reformulation of the original mode coupling
equations \eref{ModecouplingRS}.
In order to proceed we make impose successively various conditions
(Conditions 1 - 3). We stress that conditions 1 and 2 do not lead 
to any loss of generality in the subsequent treatment.\\

\noindent \underline{\bf Condition 1:} Scaling ($k\neq 0$).\\

\noindent The mode coupling equation \eref{ModecouplingFT} can be further
analyzed using the scaling form \eref{scalingformFT2}.
To this end we first rewrite \eref{ModecouplingFLT}
in terms of $\tilde{\omega}_\alpha := \omega + i v_\alpha k$. This yields
\bel{ModecouplingFLT2}
\tilde{S}_\alpha(k,\tilde{\omega}_\alpha) = \hat{S}_\alpha(k,0)
\left[\tilde{\omega}_\alpha + D_\alpha k^2
+ 2 \sum_{\beta,\gamma}
(G^{\alpha}_{\beta\gamma})^2 I_{\beta\gamma}(k,\tilde{\omega}_\alpha) \right]^{-1}
\ee
with modified memory integral
\bel{memoryFLT2}
I_{\beta\gamma}(k,\tilde{\omega}_\alpha) = k^2
\int_{0}^\infty \rmd s\, \rme^{-(\tilde{\omega}_\alpha-i v_\alpha k)  s}
\int_{-\infty}^\infty \rmd q\,\hat{S}_\beta(q,s) \hat{S}_\gamma(k-q,s).
\ee

Using the scaling ansatz \eref{scalingformFT2} we arrive at
\bea
\label{memoryFLT2b}
I_{\beta\gamma}(k,\tilde{\omega}_\alpha) & = & k^2
\int_{0}^\infty \rmd s\, \rme^{-(\tilde{\omega}_\alpha + i (v_\gamma-v_\alpha) k)  s}
A_{\beta\gamma}(k,s) \\
\label{memoryFLT2c}
& = &
 k^2  \int_{0}^\infty \rmd s\, \rme^{-(\tilde{\omega}_\alpha + i (v_\beta-v_\alpha) k)  s}
A_{\gamma\beta}(k,s)
\eea
with
\bel{memoryint2}
A_{\beta\gamma}(k,s) = \int_{-\infty}^\infty \rmd q\,
\rme^{i (v_\gamma- v_\beta) q s}
\hat{f}_\beta(qs^{\frac{1}{z_\beta}}) \hat{f}_\gamma((k-q)s^{\frac{1}{z_\gamma}}).
\ee

As pointed out above, in the static case $k=0$ the constant solution to the
mode coupling equations is exact. Therefore we can focus on the non-static case
$k\neq 0$. With $k=|k|\sign(k)$ and
the substitution of integration variables
$|k| s \to s$ we obtain
\bea
\label{memoryFLT2d}
I_{\beta\gamma}(k,\tilde{\omega}_\alpha) & = & |k|
\int_{0}^\infty \rmd s\, \rme^{-(\tilde{\omega}_\alpha |k|^{-1} + i (v_\gamma-v_\alpha) \sign(k)) s}
B_{\beta\gamma}(k,s) \\
\label{memoryFLT2e}
& = &
|k| \int_{0}^\infty \rmd s\, \rme^{-(\tilde{\omega}_\alpha |k|^{-1}+ i (v_\beta-v_\alpha)\sign(k)) s}
B_{\gamma\beta}(k,s)
\eea
with
\bel{memoryint3}
B_{\beta\gamma}(k,s) = \int_{-\infty}^\infty \rmd q\,
\rme^{i (v_\gamma- v_\beta) q |k|^{-1} s}
\hat{f}_\beta(q |k|^{-\frac{1}{z_\beta}}s^{\frac{1}{z_\beta}})
\hat{f}_\gamma((k-q)|k|^{-\frac{1}{z_\gamma}}s^{\frac{1}{z_\gamma}}).
\ee
\phantom{We}\\

\noindent \underline{\bf Condition 2:} Local interactions
($z_\alpha > 1$ $\forall \alpha$).\\

\noindent As discussed above, for sufficiently fast decaying interaction
strength one expects that all modes spread sub-ballistically around their
centers at $x_\alpha(t)$, i.e., $z_\alpha > 1$ $\forall \alpha$.  Then
the small-$k$ behaviour of the integral \eref{memoryint3} simplifies
since the term $k|k|^{-\frac{1}{z}_\gamma}$ in the second argument
vanishes. One is left with 
\bel{memoryint4} B_{\beta\gamma}(k,s) =
\int_{-\infty}^\infty \rmd q\, \rme^{i (v_\gamma- v_\beta) q |k|^{-1}
  s} \hat{f}_\beta(q |k|^{-\frac{1}{z_\beta}}s^{\frac{1}{z_\beta}})
\hat{f}_\gamma(-q|k|^{-\frac{1}{z}_\gamma}s^{\frac{1}{z}_\gamma}).
\ee

For $v_\gamma = v_\beta$ this expression reduces to
\bel{memoryint5}
B_{\beta\gamma}(k,s) = \int_{-\infty}^\infty \rmd q\,
\hat{f}_\beta(q |k|^{-\frac{1}{z_\beta}}s^{\frac{1}{z_\beta}}) 
\hat{f}_\gamma(-q|k|^{-\frac{1}{z}_\gamma}s^{\frac{1}{z}_\gamma}).
\ee
Taking $\beta=\gamma$ this yields the diagonal elements
\bea
\label{memoryint6}
B_{\beta\beta}(k,s) & = & \int_{-\infty}^\infty \rmd q\,
\hat{f}_\beta(q |k|^{-\frac{1}{z_\beta}}s^{\frac{1}{z_\beta}}) 
\hat{f}_\beta(-q|k|^{-\frac{1}{z}_\beta}s^{\frac{1}{z_\beta}})
\nonumber\\
& = & |k|^{\frac{1}{z_\beta}}s^{-\frac{1}{z_\beta}} \Omega[\hat{f}_\beta]
\eea
with the functional
\be
\Omega[f] = \int_{-\infty}^\infty \rmd k\, \hat{f}(k)\hat{f}(-k)
= \int_{-\infty}^\infty \rmd x\, (f(x))^2 .
\label{defOmega}
\ee

Thus we find from \eref{memoryFLT2d}
\be
I_{\beta\beta}(k,\tilde{\omega}_\alpha) =
|k|^{1+\frac{1}{z_\beta}} \Omega[\hat{f}_\beta] \int_{0}^\infty \rmd s\, 
s^{-\frac{1}{z_\beta}}
\rme^{-(\tilde{\omega}_\alpha |k|^{-1}+ i (v_\beta-v_\alpha)\sign(k)) s}.
\ee
With the scaling variable
\be
\zeta_\alpha = \tilde{\omega}_\alpha |k|^{-z_\alpha}
\ee
and the shorthand
\be
v_k^{\alpha\beta} = (v_\alpha-v_\beta)\, \sign(k)
\ee
this reads
\bea
I_{\beta\beta}(k,\zeta_\alpha) & = &
 |k|^{1+\frac{1}{z_\beta}} \Omega[\hat{f}_\beta] \int_{0}^\infty \rmd s \,
 \rme^{-(\zeta_\alpha|k|^{z_\alpha -1} -i v_k^{\alpha\beta}) s} 
s^{-\frac{1}{z_\beta}}\\
\label{memoryint6a}
& = &  |k|^{1+\frac{1}{z_\beta}} \Omega[\hat{f}_\beta] \Gamma
\left(1-\frac{1}{z_\beta}\right) \left(\zeta_\alpha|k|^{z_\alpha -1} 
-i v_k^{\alpha\beta}\right)^{\frac{1}{z_\beta}-1}
\eea
which also holds for $\beta=\alpha$.
Here we have used the integral representation
\be
\Gamma(x) = p^{x} \int_0^\infty \rmd u\, u^{x-1} \rme^{-pu}
= p^{x}/x \int_0^\infty \rmd u\, \rme^{-pu^{1/x}}
\label{GammaIntegral}
\ee
for $\Re(x) > 0,\,\Re(p) > 0$ of the Gamma-function.\\

\noindent \underline{\bf Condition 3:} Strict hyperbolicity
($v_\beta \neq v_\gamma$ $\forall \beta \neq \gamma$).\\

\noindent Up to this point the assumption of strict hyperbolicity
has only led us to consider the mode-coupling equations in the form
\eref{ModecouplingRS}, but it has not yet entered their analysis.
Strict hyperbolicity plays a role only in \eref{memoryint4}.
We make the substitution of integration variables
$q(s/|k|)^x \to q$ where
$x=\max{[\frac{1}{z_\beta},\frac{1}{z}_\gamma]}<1$. Then
\eref{memoryint4} becomes
\bel{memoryint4a}
B_{\beta\gamma}(k,s) = |k/s|^x \int_{-\infty}^\infty \rmd q\,
\rme^{i (v_\gamma- v_\beta) q |k/s|^{x-1}}
\hat{f}_\beta(q |k/s|^{x-\frac{1}{z_\beta}})
\hat{f}_\gamma(-q|k/s|^{x-\frac{1}{z}_\gamma}).
\ee
This leads to a term $|k|^{x-1} \to \infty$ in the exponential.
Thus for $v_\gamma \neq v_\beta$  we have a rapidly oscillating term and
the integral vanishes exponentially fast.

This proves that the leading contributions to the dynamical structure
function come from the diagonal elements $\beta=\gamma$ of the mode coupling
matrix. Therefore \eref{ModecouplingFLT2} reads
\be
\label{ModecouplingFLT3}
\tilde{S}_\alpha(k,\zeta_\alpha) = \frac{1}{\sqrt{2\pi}} |k|^{-z_\alpha}
h_\alpha(\zeta_\alpha)
\ee
where from \eref{memoryint6a} we have
\bea
h_\alpha(\zeta_\alpha) & = & \lim_{k\to 0}
 \left[\zeta_\alpha + D_\alpha |k|^{2-z_\alpha}
 + Q_{\alpha\alpha}
  \zeta_\alpha^{\frac{1}{z_\alpha}-1} |k|^{3-2 z_\alpha}
 \right. \nonumber \\
 \label{memoryLFT4}
& & \left. +   \sum_{\beta\neq \alpha} Q_{\alpha\beta}
\left(\zeta_\alpha|k|^{z_\alpha -1}-i 
v_k^{\alpha\beta}\right)^{\frac{1}{z_\beta}-1} 
|k|^{1+\frac{1}{z_\beta}-z_\alpha} 
\right]^{-1}.
\eea
with the generally positive constants
\bel{Qab}
Q_{\alpha\beta} = 2 (G^{\alpha}_{\beta\beta})^2
\Gamma\left(1-\frac{1}{z_\beta}\right)\Omega[\hat{f}_\beta] \geq 0.
\ee

We invoke again strict hyperbolicity and subballistic scaling to
deduce that the term $\zeta_\alpha|k|^{z_\alpha -1}$ in \eref{memoryLFT4}
can be neglected for the long wave length behaviour.
This yields for the diagonal terms
\bea
h(\zeta_\alpha) & = & \lim_{k\to 0}
 \left[\zeta_\alpha + D_\alpha |k|^{2-z_\alpha}
 + Q_{\alpha\alpha}
  \zeta_\alpha^{\frac{1}{z_\alpha}-1} |k|^{3-2 z_\alpha} \right. 
\nonumber \\
\label{memoryLFT5}
 & & \left.  +  \sum_{\beta\neq \alpha} Q_{\alpha\beta} 
\left(-i v_k^{\alpha\beta}\right)^{\frac{1}{z_\beta}-1}
|k|^{1+\frac{1}{z_\beta}-z_\alpha}
\right]^{-1}.
\eea
This is the starting point for the subsequent analysis of the small-$k$ 
behaviour. We remark that with the shorthand
\be
\sigma_k^{\alpha\beta} = \sign[k(v_\alpha-v_\beta)]
\ee
we have
\bea
\label{ivab1}
\left(-i v_k^{\alpha\beta}\right)^{\frac{1}{z_\beta}-1} 
& = & |v_\alpha-v_\beta|^{\frac{1}{z_\beta}-1} \exp{\left(i 
\sigma_k^{\alpha\beta} \left(1-\frac{1}{z_\beta}\right)\frac{\pi}{2}\right)}\\
\label{ivab2}
& = & \frac{\cos{\left(\left(1-\frac{1}{z_\beta}\right)\frac{\pi}{2}\right)} }
{|v_\alpha-v_\beta|^{1-\frac{1}{z_\beta}}}
\left[1 + i \sigma_k^{\alpha\beta} \tan{\left(\left(1-\frac{1}{z_\beta}
\right)\frac{\pi}{2}\right)}\right]\\
\label{ivab3}
& = & \frac{\sin{ \left(\frac{\pi}{2z_\beta}\right)  }}
{|v_\alpha-v_\beta|^{1-\frac{1}{z_\beta}}}
\left[1 - i \sigma_k^{\alpha\beta} \tan{\left(  \left(1+\frac{1}{z_\beta}
\right)\frac{\pi}{2}\right)}\right]
\eea
In the last line we made use of $\tan{(-x)}=-\tan{(x)}$ and
$\tan(x) = \tan{(x-\pi)}$.

\subsection{Asymptotic analysis}

Now one has to search for the dynamical exponents for which the limit
$k\to 0$ is non-trivial, i.e.,
$h(\zeta_\alpha)$ finite and $h(\zeta_\alpha)\neq \zeta_\alpha$ (which
would correspond to the $\delta$-peak of the linear theory which does
not exhibit the fluctuations).  This has to be done self-consistently for
all modes. Different self-consistency conditions arise depending on
which diagonal elements of the mode-coupling matrices vanish.  In the
following we consider some fixed mode $\alpha$ and study all possible
scenarios which depend on which is the smallest power in $k$ in
\eref{memoryLFT5} that yields a non-trivial scaling form.  
To this end we define the set 
\begin{equation}
\I_\alpha := \{\beta: G^{\alpha}_{\beta\beta} \neq 0\}  
\label{setdefinition}
\end{equation}
of non-zero diagonal mode coupling coefficients. Thus $\I_\alpha$ is the set of 
modes $\beta$ that give rise to a non-linear term in the time-evolution of the
mode $\alpha$ that one considers.\\ 

\noindent \underline{\bf Case A: $\I_\alpha = \emptyset$} \\

\noindent If mode $\alpha$ decouples, i.e., if {\it all} diagonal terms
$G^{\alpha}_{\beta\beta} = 0$ then
one has $h(\zeta_\alpha) =
 \left[\zeta_\alpha + D_\alpha |k|^{2-z_\alpha}
\right]^{-1}$ and therefore
\bel{dynexpdiff}
z_\alpha=2
\ee
and
\bel{scalefundiff}
\hat{S}_\alpha(k,t) = \frac{1}{\sqrt{2\pi}} \rme^{-iv_\alpha k t 
- D_\alpha k^2 t}
\ee
which is pure diffusion. (We remind the reader that we
ignore possible logarithmic corrections from cubic contributions
to the NLFH equations.)

From \eref{scalefundiff} we read off the scaling function
\be
\hat{f}_\alpha(\kappa_\alpha) = \frac{1}{\sqrt{2\pi}}
\rme^{-D_\alpha\kappa_\alpha^2}
\ee
with scaling variable $\kappa_\alpha = k t^{1/2}$.
This yields
\bel{Omegadiff}
\Omega[\hat{f}_\alpha] = \frac{1}{2\sqrt{2\pi D_\alpha}} \quad 
\mbox{ for diffusive modes } \alpha
\ee
and
\bel{Qabdiff}
Q_{\beta\alpha} = \frac{ (G^{\beta}_{\alpha\alpha})^2}{\sqrt{2 D_\alpha}} 
\quad \mbox{ for non-diffusive modes } \beta \neq \alpha .
\ee

\noindent \underline{\bf Case B: $\alpha \notin \I_\alpha, \,\I_\alpha 
\neq \emptyset$}\\

\noindent If $G^{\alpha}_{\alpha\alpha} = 0$, but some 
$G^{\alpha}_{\beta\beta} \neq 0$, then mode $\alpha$ has quadratic contributions
from one ore more other modes $\beta$. One has
\be
h(\zeta_\alpha) = \lim_{k\to 0}
 \left[\zeta_\alpha + D_\alpha |k|^{2-z_\alpha}
  +  \sum_{\beta\neq \alpha} Q_{\alpha\beta} \left(-i v_k^{\alpha\beta}\right)^{\frac{1}{z_\beta}-1}
|k|^{1+\frac{1}{z_\beta}-z_\alpha}
\right]^{-1}.
\ee
corresponding to
\be
\label{memoryFT2}
\hat{S}_\alpha(k,t) = \frac{1}{\sqrt{2\pi}} \exp{\left(-iv_\alpha k t - 
\left[ D_\alpha k^2 + \sum_{\beta} Q_{\alpha\beta} \left(-i 
v_k^{\alpha\beta}\right)^{\frac{1}{z_\beta}-1}
|k|^{1+\frac{1}{z_\beta}} \right] t\right)}
\ee
Since by Condition 2 one has $1+\frac{1}{z_\beta} <2$ it follows that 
$2-z_\alpha > 1+\frac{1}{z_\beta}-z_\alpha$. Hence the diffusive term in \eref{memoryFT2}
is subleading and the dominant terms in \eref{memoryFT2}
are those terms proportional to $(G^{\alpha}_{\beta\beta})^2$ which have 
the largest $z_\beta$.
We shall denote this value by $z_\beta^{max}$ and the define the set 
$\I_\alpha^\ast = \{\beta \in \I_\alpha: z_\beta = z_\beta^{max}\}$. 
This leads to
\bel{dynexpLevy}
z_\alpha =
\min_{\beta \in \I_\alpha}\left[\left(1 + \frac{1}{z_\beta}\right)\right] 
= 1 + \frac{1}{z_\beta^{max}} > 1.
\ee
Hence the assumption of subballistic scaling that arises from Condition 2 
is self-consistent. The dynamical structure \eref{memoryFT2} reduces to
\be
\label{scalefunLevy1}
\hat{S}_\alpha(k,t) = \frac{1}{\sqrt{2\pi}} \exp{\left(-iv_\alpha k t -
\sum_{\beta\in \I_\alpha^\ast} Q_{\alpha\beta} \left(-i v_k^{\alpha\beta}
\right)^{z_\alpha-2} |k|^{z_\alpha} t\right)}
\ee
where from \eref{ivab3} and \eref{dynexpLevy} we have
\be
\left(-i v_k^{\alpha\beta}\right)^{z_\alpha-2} =
\frac{ \sin{\left(\left(z_\alpha-1\right)\frac{\pi}{2}\right)} }{ 
|v_\alpha-v_\beta|^{2- z_\alpha} } \left(1 - i \sigma_k^{\alpha\beta} 
\tan{\left( \frac{\pi z_\alpha}{2}\right)}\right).
\ee

Defining
\bea
\label{ELevy}
E_\alpha & = & \sum_{\beta\in \I_\alpha^\ast} Q_{\alpha\beta}
\frac{ \sin{\left(\left(z_\alpha-1\right)\frac{\pi}{2}\right)} }{ 
|v_\alpha-v_\beta|^{2- z_\alpha} }\\
\label{FLevy}
F_\alpha & = & \sum_{\beta\in \I_\alpha^\ast} Q_{\alpha\beta}
\frac{ \sin{\left(\left(z_\alpha-1\right)\frac{\pi}{2}\right)} }{ 
|v_\alpha-v_\beta|^{2- z_\alpha} }
\sign{(v_\alpha-v_\beta)}\\
\label{ALevy}
A_\alpha & = & \frac{F_\alpha}{E_\alpha}
\eea
allows us to write
\be
\label{scalefunLevy}
\hat{S}_\alpha(k,t) = \frac{1}{\sqrt{2\pi}} \exp{\left(-iv_\alpha k t 
- E_\alpha |k|^{z_\alpha} t \left[1- i A_\alpha \tan{\left(
\frac{\pi z_\alpha }{2}\right)} \sign{(k)} \right]\right)}.
\ee
One recognizes in \eref{scalefunLevy} an asymmetric $\alpha$-stable
L\'evy-distribution with asymmetry $A_\alpha \in [-1,1]$.  If mode
$\alpha$ is to the left or right of {\it all} modes with
$z_\beta^{max}$ that control it (i.e.  if $v_\alpha < v_\beta \forall
\beta \in \I_\alpha^\ast$ or if $v_\alpha > v_\beta \forall \beta \in
\I_\alpha^\ast$, then $\sigma_k^{\alpha\beta}$ has the same sign for
all $\beta \in \I_\alpha^\ast$ and as a consequence $A_\alpha = \pm
1$.  This means that then the asymmetry is maximal. This is the
classical analogue of the Lieb-Robinson bound which is a theoretical
upper limit on the speed at which information can propagate in
non-relativistic quantum systems \cite{Lieb72}.

From \eref{scalefunLevy} we obtain the scaling function
\be
\hat{f}_\alpha(\kappa) = \frac{1}{\sqrt{2\pi}} \exp{\left(- E_\alpha 
|\kappa|^{z_\alpha} \left[1- i A_\alpha \sign{(\kappa)} 
\tan{\left(\frac{\pi z_\alpha }{2}\right)}\right]
\right)}
\ee
which gives (see (\ref{defOmega}) and (\ref{GammaIntegral}))
\bel{OmegaLevy}
\Omega[\hat{f}_\alpha] = \frac{1}{\pi z_\alpha}\left(
2E_\alpha\right)^{-\frac{1}{z_\alpha}}
\Gamma\left(\frac{1}{z_\alpha}\right) \quad \mbox{ for Fibonacci modes }
\alpha.
\ee
Using the identity
\bel{Gammaid}
\Gamma\left(1-\frac{1}{x}\right) = \frac{\pi}{\Gamma\left(
\frac{1}{x}\right)\sin{\left(\frac{\pi}{x}\right)}}
\ee
one finds
\bel{QabLevy}
Q_{\alpha\beta} =
\frac{2(G^{\alpha}_{\beta\beta})^2 \left(2E_\beta\right)^{-\frac{1}{z_\beta}}}
{z_\beta \sin{\left(\frac{\pi}{z_\beta}\right)}}
 \quad \mbox{ for Fibonacci modes } \beta \neq \alpha
\ee
for the constant \eref{Qab}.
We recall that $E_\alpha$ is not a simple constant depending only on
mode $\alpha$, but a functional that depends on all modes $\beta \in
\I_\alpha^\ast$.\\ 

\noindent The upshot of cases A and B is that if $G^{\alpha}_{\alpha\alpha} 
= 0$ one has the bounds 
\bel{dynexpbounds}
1< z_\alpha \leq 2
\ee
for the dynamical exponents of modes whose self-coupling constant vanishes.
The equality $z=2$ is attained if and only if all diagonal coupling constants 
of that mode vanish.
The relation (\ref{dynexpLevy}) determines the dynamical exponents.
The scaling functions are asymmetric L\'evy functions.\\

\noindent \underline{\bf Case C: $\alpha \in \I_\alpha$}\\

\noindent For $G^{\alpha}_{\alpha\alpha} \neq 0$, i.e., non-vanishing
quadratic self-coupling, imagine first that $z_{\beta^\star} >2$ 
for some mode $\beta^\star$. 
Then according to \eref{memoryLFT5} a non-trivial
scaling form is obtained for the following values of the dynamical exponent:
$z_\alpha = 1 + 1/z_{\beta^\star} < 3/2$ (from the term
proportional to $G^{\alpha}_{{\beta^\star}{\beta^\star}}$), $z_\alpha = 3/2$
(from the self-coupling term  $G^{\alpha}_{\alpha\alpha}$), or $z_\alpha = 2$
(from the diffusive term). This excludes the possiblity $z_\alpha >2$ for
$G^{\alpha}_{\alpha\alpha} \neq 0$. Above it was established that
$z_\alpha \leq 2$ for $G^{\alpha}_{\alpha\alpha} = 0$. Thus we conclude that
for {\it all} modes the bounds \eref{dynexpbounds} are valid self-consistently.
Therefore below we can assume without loss of generality $1< z_\beta \leq 2$.

Next we observe the leading small-$k$ behaviour of \eref{memoryLFT5} with 
non-trivial scaling form is obtained for 
$z_\alpha = \min{\left\{ 2, 3/2, 1+1/z_\beta\right\}}$. Thus
\bel{dynexpKPZ}
z_\alpha = 3/2
\ee
because of \eref{dynexpbounds}.

Even though the dynamical exponent is uniquely given by $z_\alpha =
3/2$ if $G^{\alpha}_{\alpha\alpha} \neq 0$, there are two different
families of scaling functions. If $z_\beta < 2$ for all modes, i.e.,
if all modes have at least one non-zero diagonal element, then
\be
\label{scalefunKPZ}
h(\zeta_\alpha) = \left[\zeta_\alpha + Q_{\alpha\alpha} 
\zeta_\alpha^{-\frac{1}{3}}  \right]^{-1}.
\ee
This corresponds to the usual KPZ-mode where mode-coupling theory is known 
to be quantitative quite good but 
not exact \cite{Frey96,Cola01}. 
On the other hand, 
if $z_\beta = 2$ for some diffusive modes from a 
set $B^{diff}$, then
\be
\label{scalefunKPZmod}
h(\zeta_\alpha) = \left[\zeta_\alpha + Q_{\alpha\alpha} 
\zeta_\alpha^{-\frac{1}{3}}
+ \sum_{\beta \in B^{diff}} Q_{\alpha\beta} \left(-i v_k^{\alpha\beta}
\right)^{-\frac{1}{2}} \right]^{-1}.
\ee
This corresponds to a modified KPZ-mode \cite{Spoh15} which has not 
been studied yet in detail.

The constants defined in \eref{Qab} are
\bel{QabKPZ}
Q_{\alpha\beta} = 2 (G^{\alpha}_{\beta\beta})^2
\Gamma\left(1/3\right)\Omega[\hat{f}_{\beta}] \mbox{ for $\beta=$ KPZ, KPZ'}
\ee
for a KPZ or modified KPZ mode $\beta$.
In order to compute $\Omega_{{\rm KPZ}}\equiv \Omega[\hat{f}_{{\rm KPZ}}]$ 
for $\beta={\rm KPZ}$
we use the exact scaling form
$S_{{\rm KPZ}}(x,t) = (\lambda t)^{-2/3} f_{{\rm KPZ}} 
((x-v_{\beta}t)/(\lambda t)^{2/3})$
with
$\lambda = 2\sqrt{2} |G^{\beta}_{\beta\beta}|$ \cite{Spoh14}.
With the scaling variable $\xi= (x-v_{\beta}t)/t^{2/3}$ as defined in 
\eref{scvxi} we obtain the real-space scaling function
$f_{\beta}(\xi) = \lambda^{-2/3} f_{{\rm KPZ}}(\lambda^{-2/3}\xi)$. 
Therefore, by definition we have
\bel{OKPZ}
\Omega_{{\rm KPZ}} = \int_{-\infty}^\infty \rmd \xi (f_{\beta}(\xi))^2 
= \lambda^{-2/3} \int_{-\infty}^\infty \rmd x (f_{{\rm KPZ}}(x))^2 
= \half (G^{\beta}_{\beta\beta})^{-2/3} c_{PS}.
\ee
For the universal constant 
\bel{cps}
c_{PS} := \int_{-\infty}^\infty \rmd x (f_{{\rm KPZ}}(x))^2 = 0.3898135914137278 
\ee
we do not have an expression in closed form  
but its value can be computed numerically with high precision from
the Pr\"ahofer-Spohn scaling function $f_{{\rm KPZ}}(x)$ tabulated in \cite{Prae04_data}. 
The double precision result (sixteen significant digits) shown in \eref{cps} is 
numerically exact and was
obtained from the data in \cite{Prae04_data} by trapezoidal 
integration.\footnote{Using the data tabulated in \cite{Prae04_data} one can calculate 
$f_{{\rm KPZ}}(x)$ with at least $90$ digits accuracy in the interval 
$x\in[-8.5 , 8.5]$. From this
one can achieve with trapezoidal integration a much higher accuracy of $c_{PS}$ 
than given here. Notice a small but significant numerical error of just over 10\%
in the value of $c_{PS}$ given below Eq. (10) in 
Ref. \cite{Das14}.} The scale factors $E_\alpha$ \eref{ELevy} 
that enter the scaling functions of Fibonacci modes with non-zero coupling to a 
KPZ mode are sensitive to $c_{PS}$ and therefore a precise
value is important for numerical fits. 
For the modified KPZ mode $\beta = {\rm KPZ'}$ the
functional $\Omega_{{\rm KPZ'}}$ has the same form as \eref{OKPZ}, but the
numerical value of the integral is not known since the 
scaling function $f_{{\rm KPZ'}}(x)$ for the modified KPZ mode is not known.

\subsection{Classification of universality classes}

We set out to classify the possible universality classes.
We summarize the equations that determine the dynamical exponents 
for a system with $n$ modes:
\be
\label{dynexpsumm2}
z_\alpha = \left\{ \ba{lcl}
2 & \mbox{ if } & \I_\alpha = \emptyset  \\
3/2 & \mbox{ if } & \alpha \in \I_\alpha\\
\min_{\beta \in \I_\alpha}\left[\left(1 + \frac{1}{z_\beta}\right)\right] 
& \mbox{ else } & \ea \right.
\ee
and
\be
\label{dynexpsumm1}
1 < z_\alpha \leq 2 \quad \forall \alpha
\ee

In order to solve the non-linear recursion (\ref{dynexpsumm2})
in case B we iterate the recursion to find e.g. for a five-fold 
iteration the continued fraction
\be
z_5=  1 +
  \cofrac{1}{
    \cofrac{1}{
      \cofrac{1}{\frac{1}{z_1}
  }}}.
\ee
Here the modes are ordered in such a fashion that the mode that minimizes
the exponent of mode 2 is mode 1 and so on.
The continued fraction terminates when a set $\I_\beta$ in this
iteration of \eref{dynexpsumm2}
is empty. Remarkably, for $z_1=1$ this is the well-known continued-fraction
representation of the Kepler ratios which implies that if $z_1$ is any
Kepler ratio $F_i/F_{i+1}$ then $z_n = F_{n+i-1}/F_{n+i}$ is also
a Kepler ratio. Thus for each parent critical exponent 2 or 3/2 from case A
or case C (which are both
Kepler ratios) one generates descendant dynamical exponents which are
Kepler ratios as long as the sets $\I_{\beta_i}$ are non-empty.
If the lowest set $\I_{\beta_1}$ is empty, i.e., if there is no
coupling from any mode in case B to a mode from case A or C then the
unique solution to the recursion is the golden mean $z_\alpha = \varphi$
for all modes from case B. The golden mean is defined by
\bel{Def:GM}
\varphi := \half (\sqrt{5}+1)\,.
\ee
Useful relations are
\bel{GMrels}
\varphi^{-1} = \half (\sqrt{5}-1), \quad \varphi = 1 + \varphi^{-1}, 
\quad \varphi^2 = 1 + \varphi,
\quad \varphi^{-2} = 2 - \varphi.
\ee
The numerical value is $\varphi\approx 1.618$.

\section{Examples}

Given as input parameters the diagonal mode-coupling constants
$G^{\alpha}_{\beta\beta}$, the diffusion coefficients $D_\alpha$ and
the KPZ-functionals $\Omega[\hat{f}_{KPZ}],\, \Omega[\hat{f}_{KPZ'}]$,
the explicit scaling solutions of the mode-coupling equations are
\eref{scalefundiff}, \eref{scalefunLevy}, \eref{scalefunKPZ} and
\eref{scalefunKPZmod}.  The dynamical exponents $z_\alpha$ have to be
determined self-consistently from the sets $\I_\alpha$ defined
in (\ref{setdefinition}), using
\eref{dynexpdiff}, \eref{dynexpLevy}, \eref{dynexpbounds} and
\eref{dynexpKPZ}.  The prefactors of the scaling variable $E_\alpha$
for the Fibonacci 
modes are then given
by \eref{Qabdiff}, \eref{ELevy}, \eref{QabLevy}, \eref{QabKPZ}. 
The asymmetry for the Fibonacci modes is determined by \eref{ALevy}. 
We stress that no
assumptions other than strict hyperbolicity and subballistic scaling
have been made to arrive at these results.  \\ 

\subsection{Example 1: $G^{1}_{11} = G^{1}_{22} = G^{2}_{22} = 0$, 
$G^{2}_{11} \neq 0$}

\noindent \underline{\bf Mode 1:}\\

\noindent For mode 1 we have case A. Eq.~\eref{dynexpdiff} gives
\be z_1=2 \ee and \eref{scalefundiff} gives \be \hat{S}_1(k,t) =
\frac{1}{\sqrt{2\pi}} \rme^{-iv_1 k t - D_1 k^2 t} \ee which is
diffusion.\\ 

\noindent \underline{\bf Mode 2:}\\

\noindent For mode 2 we have case B. Since there is only one other mode, 
which has $z_1=2$, \eref{dynexpLevy} gives 
\be 
z_2=3/2.  
\ee 
From \eref{Qabdiff} we obtain 
\be 
Q_{21} = \frac{ (G^{2}_{11})^2}{\sqrt{2 D_1}}, 
\ee 
from \eref{ELevy} 
\be E_2 = Q_{21} \frac{ \cos{\left(\left(2-
        z_2\right)\frac{\pi}{2}\right)} }{ |v_2-v_1|^{2- z_2} } =
\frac{(G^{2}_{11})^2}{2\sqrt{D_1|v_2-v_1|}} 
\ee 
and from \eref{ALevy}
\be 
A_2 = \sign{(v_2-v_1)} \,.
\ee 
Since $\tan{(z_2\pi/2)}=-1$ we arrive at 
\be 
\hat{S}_2(k,t) = \frac{1}{\sqrt{2\pi}} \exp{\left(-iv_2 k t -
    E_2 |k|^{3/2} t \left(1+ i \sign{(k(v_2-v_1))} \right) \right)}
\ee 
which is in agreement with \cite{Popk15a}.

\subsection{Example 2: $G^{1}_{11} = G^{2}_{22} = 0$, $G^{1}_{22}, 
\, G^{2}_{11} \neq 0$}

For both modes we have case B. It is expedient to define
\bea
H_\alpha & := & 2 E_\alpha \\
g_1 & := & (G^1_{22})^2 \\
g_2 & := & (G^2_{11})^2 \\
\theta & := & \frac{4 \sin{\left( (1-\varphi) \frac{\pi}{2}\right)}   }
{\varphi \sin{\left(\frac{\pi}{\varphi}\right)}
|v_1-v_2|^{2-\varphi} }
\eea

For modes 1 and 2 we have from \eref{dynexpLevy}
\be
z_1 = 1 + \frac{1}{z_2}, \quad z_2 = 1 + \frac{1}{z_1}
\ee
The solution of these two equations is
\be
z_1 = z_2 = \half (1+ \sqrt{5}) = \varphi
\ee
(see also the relations (\ref{GMrels})).

From \eref{ELevy} and \eref{QabLevy} we obtain
\bea
H_1 = Q_{12} \frac{2\sin{\left( (1-\varphi) 
\frac{\pi}{2}\right)}}{|v_1-v_2|^{2-\varphi}},
& &
H_2 = Q_{21} \frac{2\sin{\left( (1-\varphi) 
\frac{\pi}{2}\right)}}{|v_2-v_1|^{2-\varphi}}\\
Q_{12} = \frac{2 g_1 H_2^{-1/\varphi}}{\varphi \sin{
\left(\frac{\pi}{\varphi}\right) }}
& &
Q_{21} = \frac{2 g_2 H_1^{-1/\varphi}}{\varphi \sin{\left(\frac{\pi}{\varphi}\right) }}.
\eea
This yields
\be
H_1 = \theta g_1 H_2^{-1/\varphi}, \quad H_2 = \theta g_2 H_1^{-1/\varphi}.
\ee
Solving for $H_1$ gives
\be
H_1^{ \varphi - \frac{1}{\varphi} } = \frac{(\theta g_1)^\varphi}{\theta g_2}.
\ee
Using the property $\varphi - 1/\varphi =1$ of the golden mean we find
\be
E_1 = \half \left( \theta^2 g_1 g_2 \right)^{\frac{\varphi-1}{2}}
\left( \frac{g_1}{g_2} \right)^{\frac{\varphi+1}{2}}
\ee

Next we use the property of the golden mean to obtain
\be
\frac{2 \sin{\left( (1-\varphi) \frac{\pi}{2}\right)}   }
{\sin{\left(\frac{\pi}{\varphi}\right)}} =
\frac{\sin{\left( (1-\varphi) \frac{\pi}{2}\right)}   }
{\sin{\left(\frac{\pi}{2\varphi}\right)} \cos{\left(\frac{\pi}{2\varphi}\right)} }
= \frac{1}{\sin{\left(\frac{\pi\varphi}{2}\right)} } .
\ee
Thus
\be
E_1 = \half \left( \frac{2 G^1_{22} G^2_{11}}
{\varphi \sin{\left(\frac{\pi\varphi}{2}\right)} |v_1-v_2|^{2-\varphi}} \right)^{\varphi-1}
\left( \frac{G^1_{22}}{G^2_{11}} \right)^{\varphi+1}.
\ee
With a similar calculation one obtains
\be
E_2 = \half \left( \frac{2 G^1_{22} G^2_{11}}
{\varphi \sin{\left(\frac{\pi\varphi}{2}\right)} |v_1-v_2|^{2-\varphi}} \right)^{\varphi-1}
\left( \frac{G^2_{11}}{G^1_{22}} \right)^{\varphi+1}.
\ee
in agreement with \cite{Popk15a} since $(2-\varphi)(1-\varphi) 
= 1 - 2/\varphi$.\\

\noindent \underline{\it Excursion:} For $\lambda:=G^1_{22}= G^2_{11}$ 
and $c=-v_1=v_2$ this case was
treated in \cite{Spoh15} in a different way. We demonstrate how the amplitude
$E:=E_1=E_2$ arises from Eqs.~(6.11), (6.12) and (6.14) in \cite{Spoh15}.
The point to prove is
\be
E=C
\ee
where $C$ is the amplitude of the scaling variable defined in the first 
line of (6.14).

{\it Proof:} We have to compute $C$ from (6.11) and (6.12). To this end 
we define
\be
\mu := 2 (4\pi\lambda)^2 a, \quad \nu 
:= \frac{1}{\gamma} \Gamma\left(\frac{1}{\gamma}\right)
\ee
with
\bel{SpohA11}
a = (4\pi c)^{-1+1/\gamma} \frac{\pi}{2 \Gamma\left(\frac{1}{\gamma}\right)
\cos{\left(\frac{\pi}{2\gamma}\right)}}
= (4\pi c)^{\gamma -2} \frac{\pi}{2 \gamma \nu
\sin{\left(\frac{\pi\gamma}{2}\right)}}
\ee
given in (6.11). In the second equality we used 
$\cos{\left(\frac{\pi}{2\gamma}\right)} =
\sin{\left(\frac{\pi\gamma}{2}\right)}$ which follows from 
$1/\gamma = \gamma-1$.

Now observe that (6.12) yields
\bel{SpohA12}
A = (\mu A)^{-1/\gamma} \nu.
\ee
Taking this to the power $\gamma$ and using $\gamma-1=1/\gamma$
yields $A^\gamma = (\mu A)^{-1} \nu^\gamma
= \mu^{-1+1/\gamma}  \nu^{\gamma-1} A^{1/\gamma}$. Since
$\gamma - 1/\gamma=1$ we arrive at
\be
A = \mu^{-1}  (\mu\nu)^{1/\gamma}
\ee
where according to \eref{SpohA11}
\be
\mu \nu = 2 (4\pi\lambda)^2  (4\pi c)^{\gamma -2} \frac{\pi}{2 \gamma
\sin{\left(\frac{\pi\gamma}{2}\right)}} =
\frac{4^\gamma \pi^{1+\gamma} \lambda^2}{ \gamma
\sin{\left(\frac{\pi\gamma}{2}\right)}} c^{\gamma -2} .
\ee
This yields
\be
(\mu \nu)^{1/\gamma}  = 4 \pi^{\gamma} \left(
\frac{ \lambda^2}{ \gamma
\sin{\left(\frac{\pi\gamma}{2}\right)}} \right)^{1/\gamma} c^{1 -2/\gamma} .
\ee

Now we note that by definition ((6.12) and first line of (6.14))
\be
C = \half \mu (2\pi)^{-\gamma} A
\ee
which gives
\be
C = \frac{1}{4} 2^{1-\gamma} \pi^{-\gamma}  (\mu\nu)^{1/\gamma} =
\frac{1}{2^{1/\gamma}} \left(
\frac{ \lambda^2}{ \gamma
\sin{\left(\frac{\pi\gamma}{2}\right)}} \right)^{1/\gamma} c^{1 -2/\gamma}
\ee
Finally we rewrite $E$
in terms of these parameters and $\varphi=\gamma$:
\bel{ESpoh}
E = \half \left( \frac{2 \lambda^2}
{\gamma \sin{\left(\frac{\pi\gamma}{2}\right)} } \right)^{1/\gamma} (2c)^{1-2/\gamma}
= \frac{1}{2^{1/\gamma}}  \left(  \frac{\lambda^2}
{\gamma \sin{\left(\frac{\pi\gamma}{2}\right)} } \right)^{1/\gamma} c^{1-2/\gamma}
\ee
which proves $C=E$. \hfill  \qed

\subsection{Example 3: Two KPZ-modes and the heat mode}

Consider three conservation laws and label the modes by 0 and $\sigma=\pm 1$. We consider
$G^\sigma_{\sigma\sigma} =  \gamma_s$, $G^0_{00}=0$ and $G^0_{11}=- G^0_{-1-1}=\gamma_h$.
Furthermore we assume $v_\sigma = \sigma v$, $v_0=0$.

In this case $\I_\sigma^\ast = \{\sigma\}$ which means that the two 
modes $\sigma=\pm 1$ are KPZ.
Following \cite{vanB12} they can be interpreted as sound modes
and mode 0 is the heat mode. For the two sound modes one has \cite{Spoh14}
\be
\phi_{\sigma}(x,t) = (\lambda_s t)^{-2/3} f_{KPZ} 
((x-\sigma vt)/(\lambda_s t)^{2/3})
\ee
with
\be
\lambda_s = 2\sqrt{2} |\gamma_s| = 2^{3/2} |\gamma_s|.
\ee
Notice that $\lambda_s^{-2/3} = 1/2 |\gamma_s|^{-2/3}$.

For the heat mode we find from \eref{QabKPZ} the constants
$Q_{01} = Q_{0-1} = 2 \gamma_h^2 \Gamma\left(1/3\right)\Omega_{KPZ}$. 
The structure of the mode-coupling matrices yields
$\I_0^\ast = \{1,-1\}$. Therefore $z_0 = 5/3$ and from \eref{ELevy} 
one has $E_0 = 2 Q_{01} \sin{(\pi/3)} v^{-1/3}$,
$F_0 = 0$. Thus \eref{ALevy} gives $A_0=0$ and
\be
\hat{S}_{0}(k,t) = \frac{1}{\sqrt{2\pi}} \exp{\left( - E_0 |k|^{5/3} t
\right)}
\ee
with
\be
E_0 = 2
\Gamma\left(\frac{1}{3}\right) \sin{\left(\frac{\pi}{3}\right)} 
\gamma_h^2 v^{-1/3}  \gamma_s^{-2/3} c_{PS}.
\ee

In order to see that this agrees with Eq. (4.12) of Ref. \cite{Spoh14}
one has to show that $E_0 = \lambda_h (2\pi)^{-5/3}$ with
\be
\lambda_h = \lambda_s^{-2/3} (G^0_{\sigma\sigma})^2 (4\pi)^2 (2\pi c)^{-1/3}
\frac{\pi}{2 \Gamma{(2/3)} \cos{(\pi/3)}} c_{PS}
\ee
and $v=c$.
Indeed, one has, using \eref{Gammaid} with $x=3/2$,
\bea
\lambda_h (2\pi)^{-5/3} & = & 4 \lambda_s^{-2/3} (G^0_{\sigma\sigma})^2  
v^{-1/3} \frac{\pi}{2 \Gamma{(2/3)} \cos{(\pi/3)}} c_{PS} \nonumber \\
& = & 4  \gamma_h^2  v^{-1/3}  \lambda_s^{-2/3}
\frac{ \Gamma{(1/3)} \sin{(2\pi/3)} }{2  \cos{(\pi/3)}} c_{PS} \nonumber \\
& = & 4  \gamma_h^2  v^{-1/3}  \lambda_s^{-2/3}
\Gamma{(1/3)} \sin{(\pi/3)}  c_{PS} \nonumber \\
& = & 2  \gamma_h^2  v^{-1/3}  \gamma_s^{-2/3}
\Gamma{(1/3)} \sin{(\pi/3)}  c_{PS} \nonumber \\
& = & E_0
\eea
which is what needed to be shown.


\section{Conclusions}

We have shown that in the scaling limit the one-loop mode-coupling
equations for the dynamical structure function for an arbitrary number
of conservation laws in the strictly hyperbolic setting can be solved
exactly.  The solution yields a discrete family of dynamical
universality classes with dynamical exponents that are the Kepler
ratios $z_i=F_{i+2}/F_{i+1}$ which are in the range $3/2 \leq z_i \leq
2$. The largest exponent $z_1=2$ corresponds to a Gaussian diffusive
mode, possibly with logarithmic corrections (that we did not
consider). The smallest exponent $z_2=3/2$ represents three distinct
universality classes with different scaling forms of the dynamical
structure function: One has the KPZ universality class with the
Pr\"ahofer-Spohn scaling function \cite{Prae02,Prae04}, a modified KPZ
universality class with unknown scaling function \cite{Spoh15}, and a
Fibonacci mode where the scaling function is given by the $3/2$-L\'evy
stable distribution \cite{Popk14,Popk15a,Spoh15}.

All higher modes $i\geq 3$ are Fibonacci modes with $z_i$-L\'evy
stable distributions as scaling functions, including the the golden
mean $z_\infty = \varphi$. In order to have a mode $i$ with dynamical
exponent $z_i$ one needs at least $i-1$ conservation laws, with the
exception of the golden mean which requires only two conservation laws
and always appears at least twice. Thus we have shown that diffusion,
KPZ, modified KPZ and the Fibonacci family provide a {\it complete}
classification of the dynamical universality classes which we expect
to be generic for one-dimensional conservative systems where the
long-time dynamics are dominated by the long-wave length behaviour of
the modes associated with the conservation laws.\\


\noindent
{\bf Acknowledgement:}\\
This work was supported by Deutsche Forschungsgemeinschaft (DFG)
under grant SCHA 636/8-2.



\begin{thebibliography}{99}

\bibitem{vanB12}
H. van Beijeren,
Exact results for anomalous transport in one-dimensional Hamiltonian systems.
Phys. Rev. Lett. {\bf 108}, 108601 (2012).

\bibitem{Popk15a}
Popkov V, Schmidt J, Sch\"utz GM (2015)
Universality classes in two-component driven diffusive systems.
J. Stat. Phys. \textbf{160}(4) 835--860.

\bibitem{Spoh15}
Spohn H, Stoltz G (2015)
Nonlinear fluctuating hydrodynamics in one dimension: The case of
two conserved fields.
J. Stat. Phys. \textbf{160}(4) 861--884 (2015).

\bibitem{Popk15b}
V. Popkov, A. Schadschneider, J. Schmidt, G.M. Sch\"utz,
Fibonacci family of dynamical universality classes,
Proc. Natl. Acad. Science (USA) \textbf{112}(41) 12645-12650 (2015).

\bibitem{Spoh14}
H. Spohn,
Nonlinear Fluctuating hydrodynamics for anharmonic chains.
J. Stat. Phys. {\bf 154}, 1191--1227 (2014).

\bibitem{Halp15}
T. Halpin-Healy, K.A. Takeuchi,
A KPZ Cocktail-Shaken, not Stirred...,
J. Stat. Phys. {\bf 160}(4), 794--814 (2015).

\bibitem{Roy15}
A. Roy, A. Dhar, O. Narayan, and S. Sabhapandit,
Tagged Particle Diffusion in One-Dimensional Systems with Hamiltonian Dynamics-II,
J. Stat. Phys. {\bf 160}(1), 73--88 (2015).

\bibitem{Delf07}
L. Delfini, S. Denisov, S. Lepri, R.Livi,  P.K. Mohanty and A. Politi,
Energy diffusion in hard-point systems.
Eur. Phys. J. Special Topics \textbf{146}, 21--35 (2007).

\bibitem{Poli11}
A. Politi, 
Heat conduction of the hard point chain at zero pressure,
J. Stat. Mech., P03028 (2011)

\bibitem{Mend13}
C.B. Mendl and H. Spohn,
Dynamic correlators of FPU chains and nonlinear fluctuating hydrodynamics.
Phys. Rev. Lett. {\bf 111}, 230601 (2013).

\bibitem{Das14}
S. G. Das, A. Dhar, K. Saito, Ch. B. Mendl, and H. Spohn,
Numerical test of hydrodynamic fluctuation theory in the Fermi-Pasta-Ulam chain.
Phys. Rev. E {\bf 90}, 012124 (2014)

\bibitem{Das01}
D. Das,  A. Basu, M. Barma, and S. Ramaswamy,
Weak and strong dynamic scaling in a one-dimensional driven
coupled-field model: Effects of kinematic waves.
Phys. Rev. E \textbf{64}, 021402 (2001).

\bibitem{Naga05}
A. Nagar, M. Barma, and S. N. Majumdar,
Passive Sliders on Fluctuating Surfaces: Strong-Clustering States.
Phys. Rev. Lett. {\bf 94}, 240601 (2005).

\bibitem{Ferr13}
P.L. Ferrari, T. Sasamoto and H. Spohn,
Coupled Kardar-Parisi-Zhang equations in one dimension.
J. Stat. Phys. {\bf 153}, 377--399 (2013).

\bibitem{Popk14}
V. Popkov, J. Schmidt, and G.M. Sch\"utz,
Superdiffusive modes in two-species driven diffusive systems.
Phys. Rev. Lett. {\bf 112}, 200602 (2014).

\bibitem{Chak16}
S. Chakraborty, S. Pal, S. Chatterjee, and M. Barma,
Large compact clusters and fast dynamics in coupled nonequilibrium systems,
Phys. Rev. E \textbf{93}, 050102(R) (2016).

\bibitem{Bern14}
C. Bernardin and P. Gon\c calves,
Anomalous fluctuations for a perturbed Hamiltonian system with exponential interactions.
Commun. Math. Phys. {\bf 325}, 291--332 (2014).

\bibitem{Komo16}
T. Komorowski1 and S. Olla,
Ballistic and superdiffusive scales in the macroscopic evolution of a 
chain of oscillators,
Nonlinearity \textbf{29}, 962--999 (2016)

\bibitem{Spoh91}
H. Spohn,
{\it Large Scale Dynamics of Interacting Particles.}
(Springer, Berlin, 1991)

\bibitem{Kipn99}
C. Kipnis and C. Landim,
{\it Scaling limits of interacting particle systems}
(Springer, Berlin, 1999)

\bibitem{Gris11}
Grisi R, Sch\"utz GM (2011)
Current symmetries for particle systems with several conservation laws.
J. Stat. Phys. \textbf{145} 1499--1512.

\bibitem{Toth03}
T\'oth B, Valk\'o B (2003)
Onsager relations and Eulerian hydrodynamic limit for systems
with several conservation laws.
J. Stat. Phys. \textbf{112} 497--521.

\bibitem{Devi92}
Devillard P, Spohn H (1992)
Universality class of interface growth with reflection symmetry.
J. Stat. Phys. \textbf{66} 1089--1099.

\bibitem{Lieb72}
E. Lieb, D. Robinson,
The finite group velocity of quantum spin systems.
Commun. Math. Phys. \textbf{28}, 251--257, (1972)




\bibitem{Frey96}
E. Frey, U.C. T\"auber, T. Hwa,
Mode-coupling and renormalization group results for the noisy Burgers equation.
Phys. Rev. E \textbf{53}, 4424--4438 (1996).

\bibitem{Cola01}
F. Colaiori and M.A. Moore, 
Numerical Solution of the Mode-Coupling Equations for the 
Kardar-Parisi-Zhang Equation in One Dimension
Phys. Rev. E \textbf{65}, 017105 (2001).

\bibitem{Prae02}
M. Pr\"ahofer and H. Spohn, in: \textit{In and Out of
Equilibrium}, edited by V. Sidoravicius, Vol. 51 of Progress in Probability
(Birkhauser, Boston, 2002).

\bibitem{Prae04}
M. Pr\"ahofer and H. Spohn,
Exact scaling functions for one-dimensional stationary KPZ growth.
J. Stat. Phys. {\bf 115}, 255--279 (2004).

\bibitem{Prae04_data}
M. Pr\"ahofer and H. Spohn, http://www-m5.ma.tum.de/KPZ

\end{thebibliography}
\end{document}